\documentclass[aps,prl,twocolumn,superscriptaddress]{revtex4}

\usepackage{amsmath}
\usepackage[]{graphicx}
\usepackage{ifthen}

\input{MyMc.tex}

\newboolean{ShowCorrections}
\setboolean{ShowCorrections}{false}

\ifthenelse{\boolean{ShowCorrections}}{%
\usepackage[normalem]{ulem}%
\newcommand{\new}[1]{\textcolor[rgb]{0,0,1}{#1}}%
\newcommand{\old}[1]{\textcolor[rgb]{1,0,0}{\sout{#1}}}%
}{%
\newcommand{\new}[1]{#1}%
\newcommand{\old}[1]{}%
}

\newcommand{\av}[2][]{\ensuremath{\left\langle#2\right\rangle_{#1}}\xspace}

\newcommand{\gxi}{\ensuremath{\av{g\xi}}}

\newcommand{\Sec}[1]{%
}
\newcommand{\Ack}{%
}

\begin{document}

\title{Josephson junction with magnetic-field tunable ground state}

\author{E. Goldobin}
\author{D. Koelle}
\author{R. Kleiner}
\affiliation{%
  Physikalisches Institut and Center for Collective Quantum Phenomena in LISA$^+$,
  Universit\"at T\"ubingen, Auf der Morgenstelle 14, D-72076 T\"ubingen, Germany
}

\author{R. G. Mints}
\affiliation{%
  The Raymond and Beverly Sackler School of Physics and Astronomy,
  Tel Aviv University, Tel Aviv 69978, Israel
}

\date{\today}

\begin{abstract}
  We consider an asymmetric 0-$\pi$ Josephson junction consisting of $0$ and $\pi$ regions of different lengths $L_0$ and $L_\pi$. \new{As predicted earlier} this system can be described by an effective sine-Gordon equation for the spatially averaged phase $\psi$ \new{so that} the effective current-phase relation of this system includes a \emph{negative} second harmonic $\propto\sin(2\psi)$. If its amplitude is large enough, the ground state of the junction is doubly degenerate $\psi=\pm\varphi$, where $\varphi$ depends on the amplitudes of the first and second harmonics. We study the behavior of such a junction in an applied magnetic field $H$ and  demonstrate that $H$ induces an additional term $\propto H \cos\psi$ in the effective current-phase relation. This results in a non-trivial ground state \emph{tunable} by magnetic field. The dependence of the critical current on $H$ allows for revealing the ground state experimentally.
\end{abstract}



\maketitle

\Sec{Introduction}
\label{Sec:Intro}

Josephson junctions (JJs) with a phase shift of $\pi$ in the ground state\cite{Bulaevskii:pi-loop} attracted a lot of interest in the recent years\cite{Baselmans:1999:SNS-pi-JJ,Ryazanov:2001:SFS-PiJJ,Kontos:2002:SIFS-PiJJ,Weides:2006:SIFS-HiJcPiJJ,vanDam:2006:QuDot:SuperCurrRev,Gumann:2007:Geometric-pi-JJ}. These JJs can be used as on-chip phase batteries for biasing various classical\cite{Ortlepp:2006:RSFQ-0-pi} and quantum\cite{Feofanov:2010:SFS:pi-qubit} circuits. This allows for removing external bias lines and reducing decoherence.
Currently, it is possible to fabricate simultaneously both 0 and $\pi$ JJs using various technologies such as superconductor-ferromagnet heterostructures\cite{Weides:2006:SIFS-0-pi} or JJs based on d-wave superconductors\cite{VanHarlingen:1995:Review,Tsuei:Review,Smilde:ZigzagPRL}.

It would be remarkable to have a JJ (a \emph{phase battery}) providing an \emph{arbitrary} phase shift $\varphi$, rather than just 0 or $\pi$. Long arrays of 0-$\pi$-0-$\pi$-$\ldots$ JJs with short segments were suggested as systems, where the $\varphi$ JJ can be realized\cite{Mints:1998:SelfGenFlux@AltJc,Mints:2002:SplinteredVortices@GB,Buzdin:2003:phi-LJJ}. Each of the segments are assumed to have a standard current-phase relation (CPR) $j_s=\pm j_0\sin\phi$. The phase $\phi$ in these systems can be written as a sum of two terms $\phi(x)=\psi+\xi(x)\sin\psi$, where $\psi$ is a constant spatially averaged phase and $|\xi(x)|\ll 1$ is alternating on a scale of the order of the segment's length. The effective CPR then reads\cite{Mints:1998:SelfGenFlux@AltJc,Mints:2002:SplinteredVortices@GB,Buzdin:2003:phi-LJJ}
\begin{equation}
  j_s=j_1\sin(\psi) + j_2 \sin(2\psi)
  , \label{Eq:CPR2}
\end{equation}
where $j_1=\av{j_c(x)}$ is the spatially averaged critical current density and $j_2<0$ is the amplitude of the effective second harmonic. \new{We note that special stochastic distributions of facets may also lead to $j_2>0$ \cite{Ilichev:1999:InhomoPos2ndHarm}.} The value of $j_2$ depends on parameters of the junctions. For $j_2<-j_1/2$ the ground state of the system is doubly degenerate with $\psi=\pm\varphi$, where
\begin{equation}
  \varphi = \arccos\left[ -{j_1}/(2j_2) \right]
  . \label{Eq:varphi}
\end{equation}
This $\varphi$ JJ\cite{Buzdin:2003:phi-LJJ} is the generalization of a $\pi$ JJ and can provide an \emph{arbitrary} phase bias $0<\varphi<\pi$. These JJs have unusual physical properties\cite{Goldobin:CPR:2ndHarm}, \eg, non quantized Josephson vortices that were predicted\cite{Mints:1998:SelfGenFlux@AltJc} and observed experimentally\cite{Mints:2002:SplinteredVortices@GB}.

The simplest system, which is suitable for the realization of a ``short'' $\varphi$ JJ, is an asymmetric (\ie, $L_0 \neq L_\pi$) 0-$\pi$ JJ with relatively short segments of length $L_0,\,L_\pi\lesssim\lambda_J$, where $\lambda_J\propto 1/\sqrt{j_0}$ is the Josephson length. Such a system can be considered as one period of an infinitely long 0-$\pi$-0-$\pi$-$\ldots$ chain. The best method to reveal the $\varphi$ JJ experimentally is to include it in a superconducting loop (thus forming a SQUID) and to measure the spontaneously generated flux. Another option is to study these JJs in a magnetic field $H$. It is worth mentioning that there is an apparent contradiction here. On one hand, the usual linear phase ansatz $\phi(x)=Hx+\phi_0$ in asymmetric 0-$\pi$ JJ results in a critical current $I_c(H)$ with a cusp-like \emph{minimum} at $H=0$, see Eq.~(11) in Ref.~\onlinecite{VanHarlingen:1995:Review} as well as Refs.~\onlinecite{Kemmler:2010:SIFS-0-pi:Ic(H)-asymm,Weides:2010:SIFS-jc1jc2:Ic(H)}. This $I_c(H)$ should be mirror symmetric with respect to the $I_c$ axis, \ie, $I_c(-H)=I_c(+H)$. On the other hand, the presence of the second harmonic in the CPR \eqref{Eq:CPR2} and the linear phase ansatz for $\psi$ leads to a Fraunhofer-like $I_{c}(H)$ with the \emph{maximum} at $H=0$\cite{Goldobin:CPR:2ndHarm}. Moreover, when $|j_2|>0.5j_1$ there are two critical currents $I_{c\pm}(H)$, corresponding to depinning of the phase from different potential wells, but both of the $I_{c\pm}(H)$ dependences have the main maximum at $H=0$.\cite{Goldobin:CPR:2ndHarm}

In this paper we study the asymmetric 0-$\pi$ JJ in a magnetic field in detail. We demonstrate that the amplitude of the second harmonic $j_2$ in Eq.~\eqref{Eq:CPR2} depends only on the JJ parameters, while the magnetic field $H$ induces an additional term $\propto H \cos\psi$. Further, we predict the dependence of $I_c(H)$ at small magnetic fields.

\Sec{Theory}
\label{Sec:Theory}

\begin{figure}[!tb]
  \includegraphics{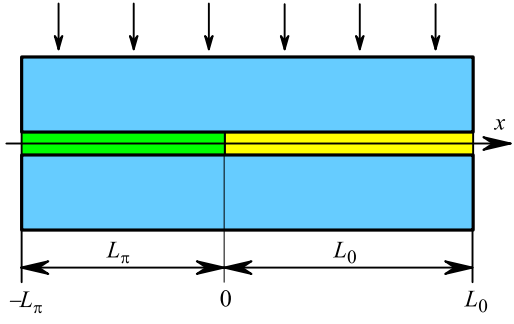}
  \caption{(Color online)
    Geometry of the 0-$\pi$ JJ.
  }
  \label{Fig:Geometry}
\end{figure}

Consider an asymmetric JJ shown in Fig.~\ref{Fig:Geometry}. The dependence $j_c(x)$ is given by
\begin{subequations}
  \begin{eqnarray}
    j_c(x)&=&-j_0,\quad \text{if }-L_\pi<x<0
    ; \label{Eq:jc(x):0}\\
    j_c(x)&=&+j_0,\quad \text{if }0<x<L_0
    \label{Eq:jc(x):pi}
  \end{eqnarray}
  \label{Eq:jc(x)}
\end{subequations}

The behavior of the phase $\phi(x)$ is defined by the static sine-Gordon equation
\begin{equation}
  \phi''(x) - j_c(x)\sin[\phi(x)] =-\gamma
  , \label{Eq:sG}
\end{equation}
where the coordinate $x$ and lengths $L_{0,\pi}$ are normalized in the usual way to $\lambda_J$ and  $\gamma=j/j_0$ is the normalized bias current density. $j_c(x)$ below is also normalized to $j_0$. Next, we write\cite{Mints:1998:SelfGenFlux@AltJc}
\begin{equation}
  j_c(x)=\av{j_c} [1+g(x)]
  , \label{Eq:g:def}
\end{equation}
where
\begin{equation}
  \av{j_c}
  = \frac{1}{L_0+L_\pi}\int_{-L_\pi}^{+L_0} j_c(x)\,dx
  = j_0\frac{L_0-L_\pi}{L_0+L_\pi}
  \label{Eq:av(j_c)}
\end{equation}
is the average value of the critical current density and
\begin{equation}
  g(x) =
  \begin{cases}
    g_\pi = \ratio{-2L_0  }{L_0-L_\pi},\text{ for } x<0;\\
    g_0   = \ratio{+2L_\pi}{L_0-L_\pi},\text{ for } x>0,
  \end{cases}
  \label{Eq:g(x)}
\end{equation}
is the deviation from the average value ($\av{g(x)}=0$).

Thus, we rewrite Eq.~(\ref{Eq:sG}) in terms of $g(x)$ and arrive to
\begin{equation}
  \phi'' - \av{j_c}[1+g(x)]\sin\phi =-\gamma
  . \label{Eq:sG(g)}
\end{equation}
In what follows we treat short JJs ($L_0,L_\pi \lesssim 1$). Therefore, the solution $\phi(x)$ can be sought in the form\cite{Mints:1998:SelfGenFlux@AltJc}
\begin{equation}
  \phi(x)=\psi + \xi(x)\sin\psi
  , \label{Eq:Ansatz}
\end{equation}
where $\psi$ is a constant\footnote{In Ref.~\onlinecite{Mints:1998:SelfGenFlux@AltJc} $\psi$ is a slowly changing function of $x$. Here, we have only one period (two segments) taken separately and assume $\psi=const$. The assumption $\psi=const$ makes the linear phase ansatz $\psi=hx+\psi_0$ used in Ref.~\onlinecite{Goldobin:CPR:2ndHarm} invalid. Instead, $h$ enters in the boundary conditions \eqref{Eq:BC:Edges} for $\xi(x)$.} and $\xi(x)$ describes small variations of the phase around $\psi$, \ie, $|\xi(x)|\ll 1$, $\av{\xi(x)}=0$.
Substituting the ansatz \eqref{Eq:Ansatz} into Eq.~\eqref{Eq:sG(g)} and expanding to the first order in $\xi(x)$ we get
\begin{equation}
  \xi'' \sin\psi - \av{j_c}[1+g(x)][1 + \xi(x)\cos\psi]\sin \psi = -\gamma
  . \label{Eq:sG:approx}
\end{equation}
Equation \eqref{Eq:sG:approx} has two types of terms: the constant ones and varying ones. Note, that the term $g(x)\xi(x)$ has both the constant (average) part $\av{g(x)\xi(x)}$ and the deviation from the average $g(x)\xi(x)-\av{g(x)\xi(x)}$.
%
%
Thus, from Eq.~\eqref{Eq:sG:approx} the relation for the constant terms reads
\begin{equation}
  \gamma = \av{j_c}\left[ \sin \psi + \av{g(x)\xi(x)} \sin \psi \cos \psi \right]
  . \label{Eq:Slow}
\end{equation}
The equation for $\xi(x)$, recalling definition \eqref{Eq:g:def}, is
\begin{equation}
  \xi'' - j_c(x)\cos\psi \xi(x)
  = \av{j_c}\left[ g(x)-\gxi\cos\psi \right]
  , \label{Eq:Fast:2RHS}
\end{equation}

It turns out that both terms $\propto \cos\psi$ have an extremely weak influence on the results. Therefore, for the sake of simplicity, we omit them right away to arrive to
\begin{equation}
  \xi'' = \av{j_c} g(x)
  . \label{Eq:Fast:Simple}
\end{equation}
Solutions of this equation include four constants to be determined from the matching conditions at $x=0$
%
\begin{equation}
  \xi_\pi(0) = \xi_0(0)
  ;\quad
  \xi'_\pi(0) = \xi'_0(0)
  , \label{Eq:BC:Zero}
\end{equation}
and boundary conditions at $x=-L_\pi$ and $x=L_0$
%
\begin{equation}
  \xi'_\pi(-L_\pi) \sin\psi = h
  ; \quad
  \xi'_0(L_0) \sin\psi = h
  , \label{Eq:BC:Edges}
\end{equation}
where $h=2H/H_{c1}$ is the normalized applied magnetic field,  $H_{c1} = \Phi_0/(\pi \lambda_J \Lambda)$ is the penetration field and $\Lambda\approx 2\lambda_L$ is the effective magnetic thickness of the JJ. Thus, we arrive to the expression for $\xi(x)$ and can calculate
\begin{equation}
  \gxi = \Gamma_0 + \Gamma_h \frac{h}{\sin\psi}
  , \label{Eq:gxi}
\end{equation}
where the coefficients
\begin{equation}
  \Gamma_0 = -\frac43 \frac{L_0^2 L_\pi^2}{L_0^2 - L_\pi^2}
  ;\quad
  \Gamma_h = \frac{L_0 L_\pi}{L_0 - L_\pi}
  . \label{Eq:Gammas}
\end{equation}
%
%
%

\begin{figure}[!tb]
  \begin{center}
    \includegraphics{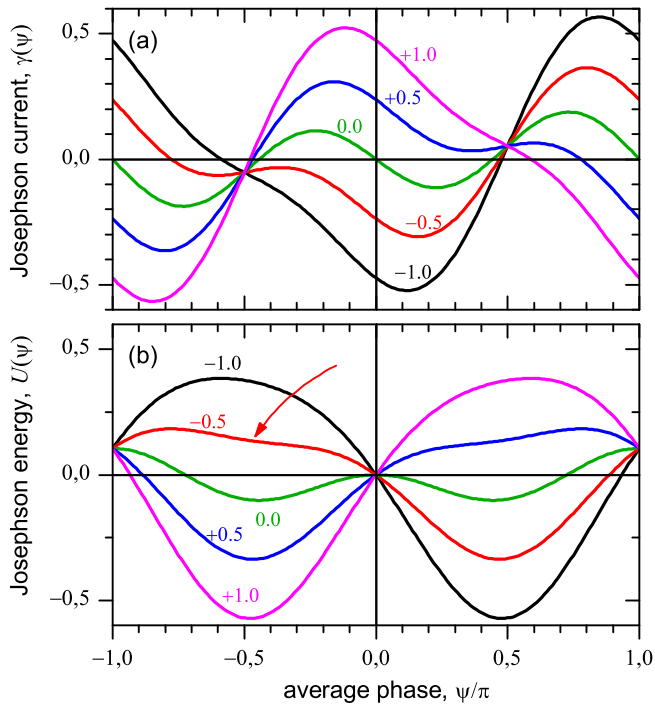}
  \end{center}
  \caption{(Color online)
    (a) effective CPR $\gamma(\psi)$ and (b) effective Josephson energy $U(\psi)$ of a 0-$\pi$ JJ with $L_0=1$, $L_\pi=0.9$ for several different values of magnetic field $h$ given next to each curve. \new{The arrow in (b) points to a local energy minimum, which appears at certain values of bias current.}
  }
  \label{Fig:CPR}
\end{figure}

Thus, Eq.~\eqref{Eq:Slow} gives the effective CPR of the JJ
\begin{equation}
  \gamma = \av{j_c}\left[
    \sin \psi + \Gamma_h h \cos\psi + \frac{\Gamma_0}{2}\sin(2\psi)
  \right]
  . \label{Eq:CPR}
\end{equation}
The result is remarkable --- as in earlier works\cite{Mints:1998:SelfGenFlux@AltJc,Mints:2002:SplinteredVortices@GB,Buzdin:2003:phi-LJJ} one gets the second harmonic $\sin(2\psi)$ with the negative amplitude of $\Gamma_0/2$. In addition, the magnetic field results in a $\Gamma_h h \cos\psi$ term which additionally modifies the CPR and is tunable by magnetic field. Fig.~\ref{Fig:CPR}(a) shows the effective CPRs $\gamma(\psi)$ of a 0-$\pi$ JJ for several different values of magnetic field $h$. At $h=0$ we might have a doubly degenerate ground state $\psi=\pm\varphi$ which, upon application of $h$, transforms into a single one.

The Josephson energy corresponding to CPR \eqref{Eq:CPR} is
\begin{equation}
  U(\psi) = \av{j_c}\left[
    1-\cos \psi + \Gamma_h h \sin\psi + \frac{\Gamma_0}{2} \sin^2 \psi
  \right]
  . \label{Eq:U(psi)}
\end{equation}
The plots of $U(\psi)$ for the same set of $h$ values is shown in Fig.~\ref{Fig:CPR}(b). Following the evolution of the curves at different $h$ one can see how the doubly degenerate ground state transforms into a single one. Moreover, one can see that at non-zero $h$ the potential energy $U(\psi)$ lacks reflection symmetry and therefore can be used to build Josephson phase ratchets\cite{Reimann:2002:BrownianMotors,Haenggi:2009:ArtBrownMotors}.

\begin{figure}[!tb]
  \begin{center}
    \includegraphics{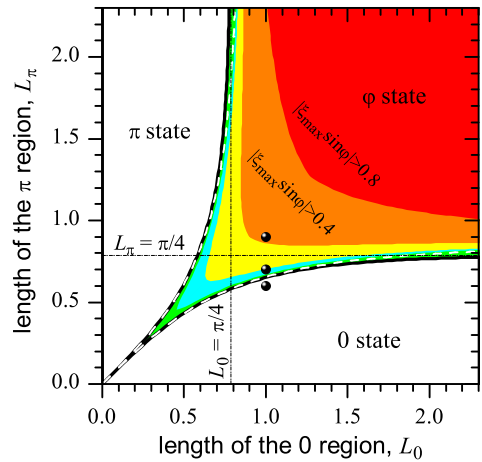}
  \end{center}
  \caption{%
    The domain of existence of a $\varphi$ state ($h=0$) on the $L_0,L_\pi$ plane. The exact boundary \eqref{Eq:Bulaevskii:varphi-domain} (continuous lines) and our approximation \eqref{Eq:Mints:varphi-domain} (dashed lines) agree very well. The $\varphi$ domain has regions of different colors from green (approximation $|\xi(x)\sin\psi|\ll 1$ works well) to red (approximation is invalid). The boundaries correspond to $|\xi_\mathrm{max}\sin\varphi|=0.1,\,0.2,\,0.4,\,0.8$. The dots indicate parameters used for discussion of Fig.~\ref{Fig:Ic(H)}. Vertical and horizontal dash-dotted lines show asymptotic behavior of exact boundary \eqref{Eq:Bulaevskii:varphi-domain}.
  }
  \label{Fig:domain}
\end{figure}

The effective CPR \eqref{Eq:CPR} at $h=0$ allows to calculate the domain of existence of nontrivial solutions, which is defined as $|\Gamma_0|>1$ \cite{Goldobin:CPR:2ndHarm}. This results in
%
  \begin{equation}
    L_\pi \geq L_0\sqrt{\ratio{3}{4L_0^2+3}}
    ;\quad 
    L_0   \geq L_\pi\sqrt{\ratio{3}{4L_\pi^2+3}}
    . 
    \label{Eq:Mints:varphi-domain}
  \end{equation}
%
Since we have made certain approximations, namely used Eq.~\eqref{Eq:Fast:Simple} instead of Eq.~\eqref{Eq:Fast:2RHS}, it is worth to compare the domain \eqref{Eq:Mints:varphi-domain} with the exact result\cite{Bulaevskii:0-pi-LJJ}, which  reads
\begin{subequations}
  \begin{eqnarray}
    L_\pi &\geq& \arctan(\tanh(L_0))
    ; \label{Eq:Bulaevskii:0-varphi}\\
    L_0   &\geq& \arctan(\tanh(L_\pi))
    , \label{Eq:Bulaevskii:pi-varphi}
  \end{eqnarray}
  \label{Eq:Bulaevskii:varphi-domain}
\end{subequations}
for our case $j_c^0=j_c^\pi$. Both boundaries, the exact one given by Eq.~\eqref{Eq:Bulaevskii:varphi-domain} and the approximate one given by Eq.~\eqref{Eq:Mints:varphi-domain} are shown in Fig.~\ref{Fig:domain}. It is seen that our approximation works extremely well not only for small $L_0$ and $L_\pi$, but also in the limits $L_0, L_\pi \to \infty$, where it deviates from the asymptotic value by only $\sim 10\units{\%}$.

\begin{figure*}[!tb]
  \begin{center}
    \includegraphics{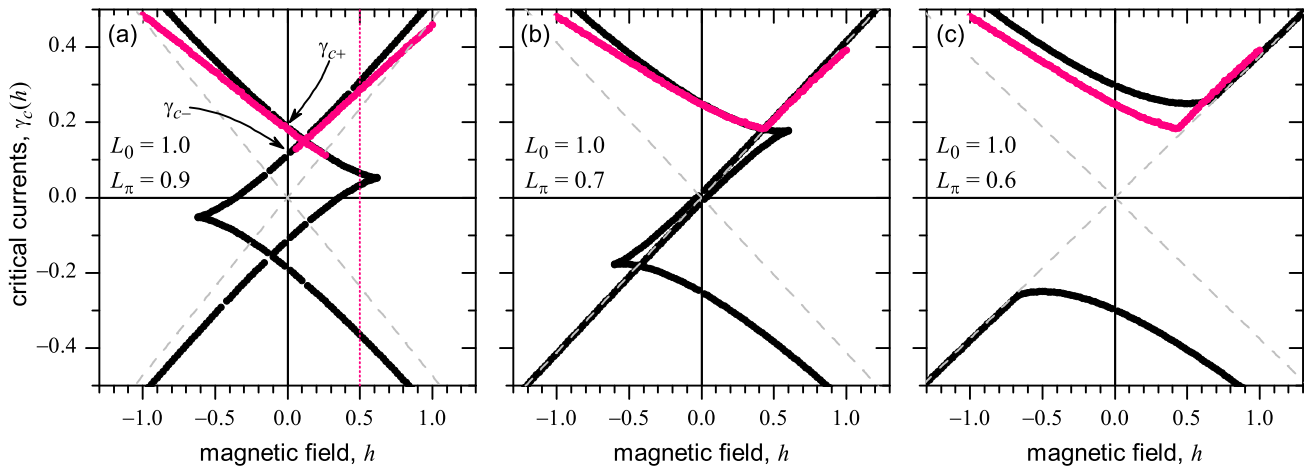}
  \end{center}
  \caption{(Color online)
    The critical current $\gamma_{c}(h)$ for 0-$\pi$ JJs with different segment lengths. Dashed lines show the asymptotic behavior \eqref{Eq:ass} for large $|h|$. Gray (pink) \new{solid} lines show the results of direct numerical simulations.
  }
  \label{Fig:Ic(H)}
\end{figure*}

In the JJ with the CPR \eqref{Eq:CPR} and energy \eqref{Eq:U(psi)}, one may have several stable static solutions and several critical (depinning) currents corresponding to the escape of the phase from the relevant energy minimum. To find the critical currents for a given magnetic field, we look for extrema of the CPR \eqref{Eq:CPR} with respect to $\psi$. We arrive to
\begin{equation}
  \cos\psi - \Gamma_h h \sin\psi - \Gamma_0 (2\cos^2\psi-1)=0
  . \label{Eq:psi_c}
\end{equation}
This problem can be reduced to a solution of a fourth order polynomial, \ie, all solutions of Eq.~\eqref{Eq:psi_c} can be found numerically. As a result we obtain up to four relevant roots and up to four corresponding critical currents.

Several examples of $\gamma_c(h)$ dependences are shown in Fig.~\ref{Fig:Ic(H)}.
One can see that for the parameter set (a), which corresponds to a state deep in the $\varphi$ region at $h=0$, see Fig.~\ref{Fig:domain}, one observes a characteristic rotated diamond-like shape with four critical currents in total for $|h|\lesssim0.6$. \old{The higher (by absolute value) critical currents correspond to the escape of the phase from the main energy minimum over the large barrier, see Fig.~\ref{Fig:CPR}(b). The lower critical currents correspond to the escape from a smaller energy minimum.}
\new{The two branches that meet at $h\approx+0.6$ ($-0.6$) correspond to the escape from the left (right) potential well in Fig.~\ref{Fig:CPR}(b). The upper (lower) branch corresponds to the escape in the right (left) direction. At $|h|\approx0.6$ the branches meet indicating the disappearance of the corresponding local energy minimum.} The two critical currents $\gamma_{c\pm}$ (in each direction) observed at $h=0$ coinside with $\gamma_{\pm}$ obtained in ref.~\onlinecite{Goldobin:CPR:2ndHarm}. One may be able to observe the lower one experimentally in a system with low enough damping using a special sweep sequence\cite{Goldobin:CPR:2ndHarm}. For some values of $h$ the smaller minimum can even be absent at $\gamma=0$ [see the arrow Fig.~\ref{Fig:CPR}(b)] and appear at larger $|\gamma|$ [the vertical dotted line corresponding to $h=0.5$ in Fig.~\ref{Fig:Ic(H)}(a)]. The small energy minimum appears for $0.03<\gamma<0.06$, which also be seen in Fig.~\ref{Fig:CPR}(b). Another distinct feature, which might be measurable in experiment, is the shift of the main minimum to $h=\pm0.11$ in Fig.~\ref{Fig:Ic(H)}(a). This shift is the evidence of a non-trivial phase state in the junction. Finally, at large $|h|$ the upper branches of $\gamma_c(h)$ approach the asymptotic lines
\begin{equation}
  \gamma_c^\mathrm{as}(h) \approx \pm \av{j_c} \Gamma_h h
  , \label{Eq:ass}
\end{equation}
which are also shown in Fig.~\ref{Fig:Ic(H)}. We remind that our analysis is valid only for small values of $|h|$ and does not reproduce the global $\gamma_c(h)$ features, presented elsewhere.

When the 0-$\pi$ JJ gets more asymmetric, see Fig.~\ref{Fig:Ic(H)}(b) and (c), the lower $\gamma_c(h)$ diamond-shaped domain becomes thinner and finally collapses. The point of collapse corresponds to the crossing of the $\varphi$-domain boundary in Fig.~\ref{Fig:domain}. Looking at $\gamma_c(h)$ one can see that the main cusp-like minimum shifts with asymmetry which can be used to extract the asymmetry from experimental data.

We also have performed numerical simulations of the $\gamma_c(h)$ dependence using \textsc{StkJJ} software\new{\cite{StkJJ}}. The above analytical results coincide with the results of direct numerical simulation\new{, shown by pink (gray) in Fig.~\ref{Fig:Ic(H)},} with good accuracy. \new{This is especially so} at $h \ll 1$ \new{when the approximation $|\xi(x)\sin\psi|\ll 1$, used to derive \eqref{Eq:sG:approx}, is valid.}

\Sec{Conclusions}
\label{Sec:Conclusions}

To summarize, we have shown that the effective CPR \eqref{Eq:CPR} includes the term $\propto H\cos\psi$, which allows to tune the ground state. Thus, our system  is a hybrid between $\varphi$ \cite{Buzdin:2003:phi-LJJ} and $\varphi_0$ \cite{Buzdin:2008:varphi0} JJs introduced earlier. The corresponding Josephson energy profile can be made asymmetric allowing to build Josephson phase ratchets.
We have clarified how the $I_c(H)$ dependence of asymmetric 0-$\pi$ JJ looks like for low magnetic field. It does have a minimum, but the minimum is shifted to some $\pm h_\mathrm{min}$ for positive and negative bias, respectively, so that the $I_c(H)$ curve is point symmetric.

\Ack

We acknowledge financial support by the German Israeli Foundation (Grant No. G-967-126.14/2007).

\bibliographystyle{apsprl}
\bibliography{SFS,SF,pi,LJJ,ratch,software}

\end{document}